# InP-based two-dimensional photonic crystals filled with polymers

R. van der Heijden[a)], C.F. Carlström, J.A.P. Snijders, R.W. van der Heijden,

F. Karouta, R. Nötzel and H.W.M. Salemink[b)]

*COBRA Inter-University Research Institute and Center for NanoMaterials, Eindhoven University of Technology, PO Box 513, NL-5600 MB Eindhoven, The Netherlands*

B.K.C. Kjellander

*Department of Applied Physics, Eindhoven University of Technology, PO Box 513, NL-5600 MB Eindhoven, The Netherlands*

C.W.M. Bastiaansen and D.J. Broer

*Department of Chemistry and Chemical Engineering, Eindhoven University of Technology, PO Box 513, NL-5600 MB Eindhoven, The Netherlands*

E. van der Drift

*Kavli Institute of Nanoscience, Delft University of Technology, P.O. Box 5053, NL-2600 GB Delft, The Netherlands*

**Abstract**

Polymer filling of the air holes of Indium Phosphide based two-dimensional photonic crystals is reported. After infiltration of the holes with a liquid monomer and solidification of the infill *in situ* by thermal polymerization, complete filling is proven using scanning electron microscopy. Optical transmission measurements of a filled photonic crystal structure exhibit a redshift of the air band, confirming the complete filling.

[a] Electronic mail: r.v.d.heijden@tue.nl
[b] Also at: Kavli Institute of Nanoscience Delft

Photonic crystals (PhCs) are artificial materials consisting of a periodic arrangement of low- and high dielectric materials in one, two or three dimensions that provide control of light at a wavelength scale[1]. Two-dimensional (2D) PhCs are of particular importance as they can be fabricated in semiconductors with available patterning and etching techniques. An epitaxially grown layer structure is used to achieve waveguiding in the 2D plane. The functionality of PhC structures will increase when their optical properties are tunable. An attractive possibility in this respect is to replace the low-index part (air) with a material that has a tunable refractive index. Liquid crystals (LCs) are particularly useful for such purpose because of the large polarization-dependent changes in refractive index ($\Delta n \sim 0.05$-$0.5$) that can be obtained. Since the first proposal of Busch and John[2], several tuning experiments have been performed with 3D (see Ref. 3 for an overview) and 2D LC-infiltrated photonic crystals.[4-10] LC tuning is inherently slow, typically in the order of ms, down to µs for certain ferroelectric or polymer-dispersed LCs[11]. Also, the liquid state could hinder certain applications or limit the compatibility with other processing steps. Hence other materials, notably electro-optically active or optically non-linear (NLO) polymers are highly desirable as tunable low-index material.

Apart from the active applications, even filling of the holes with a passive dielectric has many applications. The infilling of deeply etched photonic crystals is expected to reduce the out-of-plane scattering, because the diffraction angle when the light passes from the waveguiding dielectric into the holes is reduced. Furthermore, it was recently shown that selective filling opens up a way to create components like single mode waveguides, waveguide bends and crossings and splitters from "bulk" 2D PhCs[12]. Also, after planarization of infiltrated structures, further polymer processing may be employed to create additional functionality in the polymer layer(s) on top.

In this Letter we demonstrate the photonic band edge shift for a deeply etched 2D PhC by filling the air holes with a polymer. A simple filling procedure is described, consisting of infiltration with liquid monomer at room temperature and ambient atmosphere followed by thermal polymerization. The solid state of the infill allows for direct inspection of the infiltrated holes by cross-sectional scanning electron microscopy (SEM). The effect of the filling on the photonic band gap was investigated by optical transmission measurements.

Photonic crystals patterns (triangular lattice) were fabricated using 100 keV electron-beam lithography, reactive ion etching of a hard mask ($SiN_x$) and $Cl_2/O_2$-based inductively coupled plasma etching (ICP). The ICP etching process that is optimized for the hole profile yields ~3.4 μm deep holes, which are nearly cylindrical in the upper 2 μm[13]. The infiltration procedure relies on the capillary action of a liquid monomer inside the ~ 200 nm diameter air holes. The contact angle between the liquid and the solid, which characterizes the wetting behavior[14], is therefore an important parameter[15]. Before infiltration, the wetting of the InP surface by the liquid infiltrant was verified by wetting angle measurements. The infiltration tests were carried out on InP substrates.

Trimethylolpropane triacrylate (TMP-3A, Aldrich) was chosen as the infiltrant for its cross-linking efficiency and relatively low viscosity. A fraction of 0.5 wt. % azo-bisisobutyronitrile (AIBN, Fluka) is added to the TMP-3A as initiator for the thermal polymerization process. Various chemical treatments of the ICP-etched InP surface were examined to minimize the contact angle. The best result was obtained after a successive rinse with 10% hydrofluoric acid in water and propanol-2, which reduced the contact angle from 33° to < 4°.

After the surface preparation, a droplet of the liquid monomer is deposited onto the sample to cover the hole pattern, either in vacuum (20 mbar) or ambient atmosphere, at room temperature. The thermal polymerization is performed on a hotplate under $N_2$-flow, to minimize oxygen reaction with the radicals formed. To suppress formation of cracks in the polymer layer, the temperature is gradually increased from room temperature up to 50 $^{o}$C, where polymerization is allowed for 10 minutes. The temperature is further increased to 70 $^{o}$C in order to increase the mobility of the unreacted monomers and the sample was baked for 20 minutes at this temperature to complete the polymerization. The refractive index of the polymer was measured separately by spectroscopic ellipsometry on a layer that was spin-cast on a Si-substrate and subsequently polymerized. For wavelengths around 1550 nm the measured refractive index is 1.465 ± 0.005; the absorption in this wavelength region is negligible for our purpose.

The filled hole pattern is inspected with cross-sectional SEM. The sample is cleaved through the hole pattern under a small angle (~3$^{o}$) with the ΓK-axis, to ensure that the cleave intersects the holes. In Fig. 1, a SEM view of the holes is shown after infiltration in (a) vacuum (20 mbar) and (b) ambient atmosphere. Nearly complete infiltration is apparent for both conditions. The detachment of the polymer plug from the sidewall, visible for the center holes in Fig. 1a, is attributed to the polymerization shrinkage, which is expected to be 10-15 % for poly-acrylates. Some holes in Fig. 1a appear to be unfilled. It was observed that the polymer plugs are not cleaved, but remain on either side of the cleavage. These experiments were carried out on etch test samples with varying hole shapes. In all cases investigated it was observed that the polymer fills the holes down to the bottom, even for the irregular and conical hole shapes as in Fig 1.

No difference could be observed between infiltration under vacuum or at ambient conditions. Probably this results from the considerable compression of the residual air under the fluid column due to the capillary pressure (~ several bars). This compression provides the driving force for air diffusion through the monomer. It is estimated that the gas flow through the liquid infill (Eq. 10 in Ref. 16) is sufficient to degas the hole in a time < 1 s. This estimate is based on the $N_2$-permeability typical for polymers[16], whereas the permeability of the liquid monomer is expected to be even higher.

Infiltration experiments with polymers in solution were also performed, using standard e-beam resists. Solidification of these was achieved by evaporation of the solvent. Due to the small volume fraction of polymer in the solution, the hole filling was poor (~10 % filled) and irregular after evaporation of the solvent.

To optically characterize the effect of the infilling on the photonic stopgap, the transmission spectrum was measured for the ΓK-direction of a 10-period triangular lattice of air holes with an air-filling factor of 0.33. For these optical experiments, the holes were etched through an InP/InGaAsP/InP planar waveguide structure. Both the InGaAsP core and the InP upper cladding are 500 nm thick, leading to an effective refractive index of 3.25 for the transverse electric (TE) guided mode at a wavelength of 1550 nm. Ridge waveguides are fabricated in the same etching step for optical access to the photonic crystal patterns (see the insert in Fig. 2b). Finally the sample was cleaved at both sides perpendicular to the ridges to obtain end facets. The ICP-etching process used for this sample is a compromise between the hole shape and ridge waveguide shape. An InP control sample, suitable for SEM inspection, was etched in the same run as the planar waveguide sample. A cross-sectional SEM-view of the vertical hole-profile for this control sample is shown in Fig. 2a. It was verified before that hole shapes etched in waveguide material and InP are similar for the etch process used.

Optical measurements were performed with an end-fire technique[17]. Chopped optical power from a tunable (1470-1570 nm) polarization controlled diode laser is TE-polarized (E-field in the photonic crystal plane) and coupled into the ridge waveguides with a microscope objective (N.A. = 0.65). The transmission is collected by a similar objective and measured with an InGaAs photodetector using a lock-in amplifier. Lithographic tuning[18] was employed to cover the ΓK stopband of the photonic crystal. After measurement of the empty structure, the holes were infiltrated with TMP-3A under ambient conditions. A thick polymer droplet remained on the surface after polymerization; at the cleavage sides of the sample no polymer could be observed with SEM. The measured transmission spectra for the empty and filled PhCs are compared in Fig. 2b.

For the empty structure, the measured stopband runs from normalized frequency $a/\lambda = 0.22$ to $a/\lambda = 0.30$, where $a$ is the lattice constant of the PhC and $\lambda$ the wavelength. These $a/\lambda$ values agree with band structure calculations with a 2D plane wave method[19] (vertical lines in Fig. 2b). The steep fall of the transmission at the dielectric band edge (low frequency side) is more than three orders of magnitude. The transmission level inside the stopband is determined by the detection limit of the set-up, due to stray light. The air band edge (high frequency side) is not as steep as the dielectric band edge, which is attributed to out-of-plane losses[20]. At the calculated position of the air band edge, the transmission level is 10% of the maximum transmission in the air band. Around $a/\lambda = 0.37$ a pseudo-gap is observed[21]. Increasing the etch depth and improving the profile of the holes will lead to larger transmission in the air band and sharper edges of the stop bands[20, 21].

Upon infiltration, the air band edge exhibits a redshift to $a/\lambda = 0.285 \pm 0.05$, while no significant shift is observed for the dielectric band edge[4]. The shift of the air band edge is calculated as a function of the refractive index inside the hole. With this relation, an effective index $n_{hole}$ is obtained from the measured shift of the air band edge. The filling efficiency of the holes (volume of polymer plug divided by hole volume) $\eta$ is calculated via: $\eta = (n_{hole}-1)/(n_{polymer}-1)$, where $n_{polymer}$ is the refractive index of the polymer[6]. From this analysis the filling efficiency is determined to be $0.8 \leq \eta \leq 1$. This number is consistent with the complete infiltration of the holes followed by polymerization shrinkage of 10-15 %. In Fig. 2b it is also apparent, that the transmission is higher after infiltration for both the dielectric band (factor of 2) and the air band (factor of 3). It is verified, that the transmission increase of the ridge waveguides, which are also largely covered by the polymer, is less than 30 %. The reproducibility of the measured signal intensity is better than 30 % for the dielectric band and better than 50 % for the air band. These results suggest that the infill reduces the out-of-plane losses, particularly in the air band of the PhC. This effect is larger for the air band as this suffers more from out-of-plane losses than the dielectric band[20].

In summary we have presented a simple scheme for polymer filling of photonic crystal holes by infiltration with a liquid monomer under ambient conditions and subsequent thermal polymerization. Complete filling was proven by SEM inspection and confirmed by optical transmission measurements. This simple method should also be useful with optically active polymers to provide electro-optical tunability of the PhC structures. When applied to selective filling of particular holes, it opens the way for PhC components with adjustable or switchable optical properties.

The authors would like to thank P. Nouwens, R. van Veldhoven, E.J. Geluk, T. de Vries, M. Sander, E. Smalbrugge and H. Thijs for contributions to the experimental work and J. van der Tol, A. Kok, S. Oei, J. Haverkort and A. Silov for helpful discussions. Part of this research is supported by NanoNed, a technology programme of the Dutch ministry of Economic Affairs via the foundation STW.

**Figure captions**

**Fig. 1:** SEM view of the photonic crystal holes after infiltration (a) in vacuum and (b) ambient atmosphere with TMP-3A. These holes are etched in InP substrate with a non-optimized ICP-process that results in a tapered profile. The thick polymer layer on top is torn off during cleaving for the sample shown in (a) while it is still visible in (b). The black arrows in (a) indicate where the detachment of the polymer from the sidewall is well visible.

**Fig. 2:** (Color online) (a) SEM view of the hole profile of an InP sample etched in the same run as the sample that was optically characterized. The dashed lines indicate the location of the InGaAsP core layer in the case of a waveguiding sample. (b). Measured transmission spectrum of an empty and infiltrated triangular lattice 2D photonic crystal with 10 periods in the ΓK-direction obtained by lithographic tuning (air filling factor = 0.33). The vertical lines represent the calculated position of the band edges for both empty and fully infiltrated holes. The inset shows a SEM top view of the measured structure. For every pattern the lattice constant in nm and the covered a/λ-range is indicated in the top of the graph.

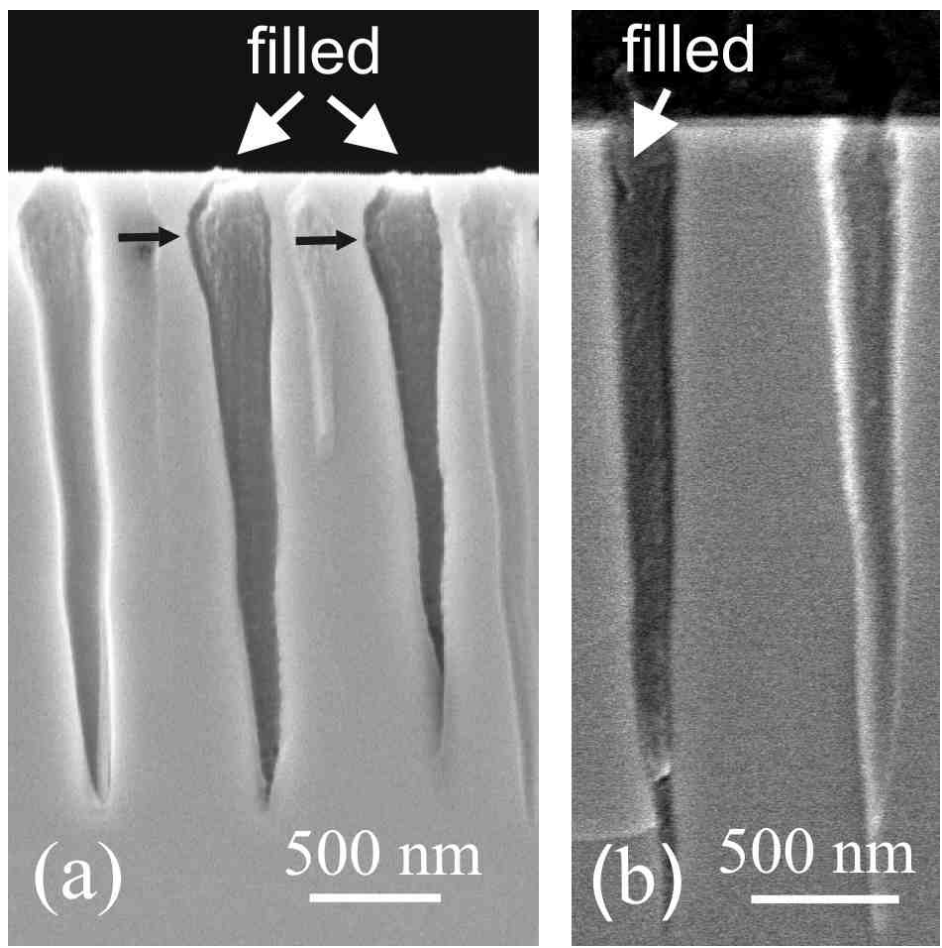

Figure 1

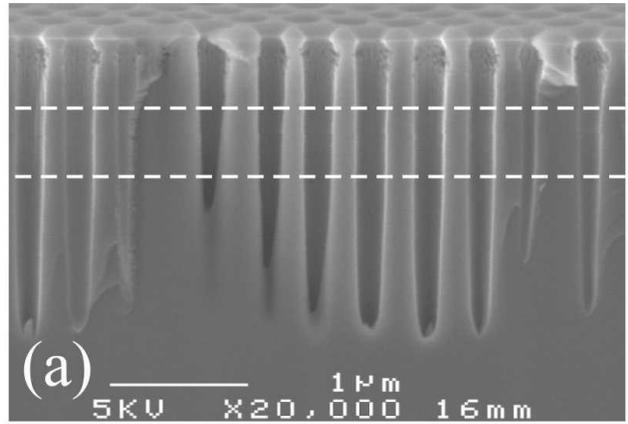

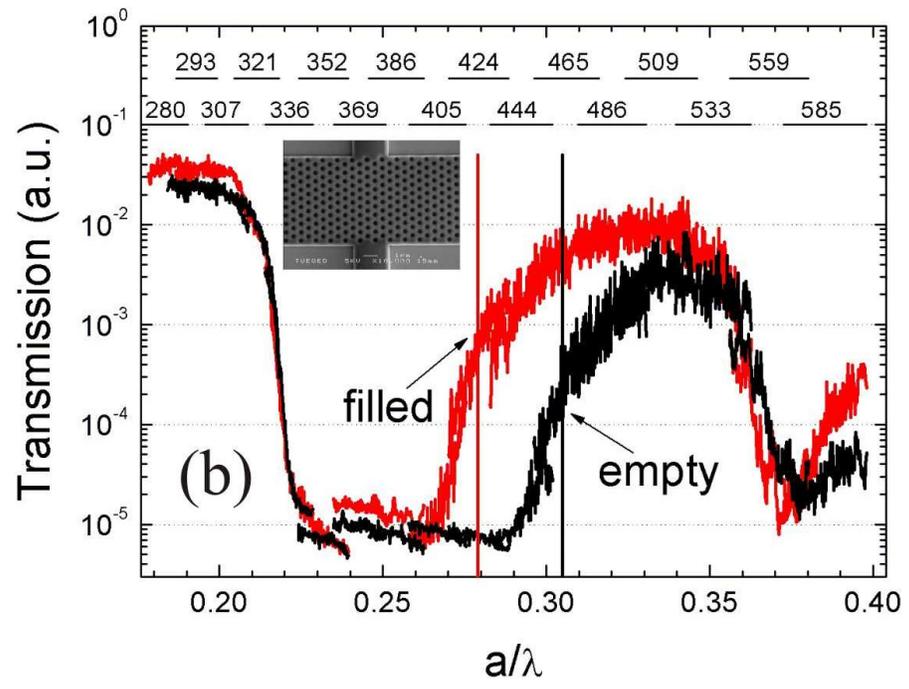

Figure 2